# Detecting Propaganda on the Sentence Level during the COVID-19 Pandemic

Rong-Ching Chang[1], Chu-Hsing Lin[2]
Department of Computer Science, Tunghai University [1, 2]
{ g08350003[1]; chlin [2] }@thu.edu.tw

**Abstract**

The spread of misinformation, conspiracy, and questionable content and information manipulation by foreign adversaries on social media has surged along with the COVID-19 pandemic. Such malicious cyber-enabled actions may cause increasing social polarization, health crises, and property loss. In this paper, using fine-tuned contextualized embedding trained on Reddit, we tackle the detection of the propaganda of such user accounts and their targeted issues on Twitter during March 2020 when the COVID-19 epidemic became recognized as a pandemic. Our result shows that the pro-China group appeared to be tweeting 35 to 115 times more than the neutral group. At the same time, neutral groups were tweeting more positive-attitude content and voicing alarm for the COVID-19 situation. The pro-China group was also using more call-for-action words on political issues not necessarily China-related.

## 1. Introduction

The novel Coronavirus (COVID-19) is a viral, respiratory disease that has spread throughout the world. The World Health Organization announced the COVID-19 outbreak to be a pandemic on March 12, 2020. To date, more than eighteen million confirmed cases have been reported, with a death toll of more than seventy thousand. Many public health policies have been put into place, such as self-quarantine, stay-at-home orders, travel-bans, and social distancing measures.

In this crisis, various research has raised concerns about the massive infodemic, a flood of false or misleading information about COVID-19 on social media platforms. The unreliable, low-quality information and conspiracy theories continuously stimulate irrational social behavior and threaten public health [19]. Research has shown that COVID-19 contributed to the rise of sinophobic content in a cross-platform manner, fueling racial discrimination [17]. At the same time, Twitter announced its Monetizable Daily Active Users (mDAU) had reached 166 million, having grown 24% during the first quarter of 2020, driven by the pandemic.

Linguistically, the defining pragmatic feature of propaganda is that it has the purpose of influencing opinion. Computational propaganda is "*propaganda created or disseminated using computational (technical) means*" [8]. Leveraging different types of social media accounts in a coordinated fashion, rhetorical, psychological propaganda techniques often convey disinformation, which is (i) false information, (ii) carrying the intention to harm. Detecting social media bots and trolls is increasingly challenging because they evolve with AI technologies that generate human-like text content with GPT-2 [1], profile pictures with DeepFake [20], audios, videos, and even fake profiles in professional network platforms such as LinkedIn. Several studies have concentrated on using supervised or semi-supervised machine learning methods. Such approaches depend heavily on the domain-driven labeling of training data, which may be labor-intensive [23][15] or vulnerable to the training set [10]. Unsupervised classification has shown high accuracy only on vocal Twitter users [15].

Research, moreover, shows that social media has become a tool for modern information-age warfare [12]. An information war is an operation to gain information advantages over an opponent. For example, it has been shown that China's coronavirus disinformation campaigns are integral to its global information warfare strategy [18]. The Chinese Communist Party, CCP, was found to be exploiting proxy accounts and bots to disseminate false stories on numerous social media platforms globally. Chinese state-linked action and coordination of Twitter accounts spread pro-Chinese government narratives as emerging events -- COVID-19 [14]. Chinese-run accounts were showing 6% more positive content, promoting the effectiveness of the government in combatting the Coronavirus; meanwhile in the Western media, the strength of positive and negative sentiment has been more balanced [9]. Chinese state-sponsored influence campaigns were found to be promoting misleadingly reframed events, amplifying conspiracy theories, and praising the Chinese government over the handling of the COVID-19 epidemic [19].

To further investigate this situation, in our research, we focused on identifying Twitter accounts, which were spreading narratives favoring CCP propaganda, labelled "pro-China" in the rest of the study. In the proposed framework, we fine-tuned the detection of propaganda at the sentence-level using contextualized word embeddings with cross-platform features from Reddit. We used this structure to predict the unlabeled tweets from the Twitter dataset. Consequently, we were able to predict at the first-level of classification whether tweets were pro-China or not.

We chose Reddit as our base dataset because of Reddit's subreddit community structure. That is, Reddit has an official COVID-19 community with the name /r/Coronavirus. It is a team of more than sixty content moderators, consisting of infectious disease researchers, virologists, computer scientists, doctors, and nurses, who spend hours daily to maintain and remove misinformation, trolls, and discussions in the threads. Additionally, there are other subreddits such as /r/China_flu, which tend to exhibit more racist behavior [13] and /r/SINO, which is mostly promoting the CCP and criticizing foreign adversaries.

## 2. Related Work

### 2.1 Contextualized Embeddings

Bidirectional Encoder Representations from Transformers (BERT) is a general pre-trained language model. It was pre-trained on a Wikipedia text set. BERT has shown promising results on stance detection tasks and has been widely used in the detection of misinformation during COVID-19. RoBERTa [24] stands for A Robustly Optimized BERT Pretraining Approach. RoBERTa is an optimized version of BERT.

### 2.2 Botometer

A Bot (also known as Sybil account) is a social media account controlled partially or entirely by software [7]. There is a wide range of the use of bots: they can be helpful in cases such as reporting weather or news, but they can be harmful when they coordinate to push or amplify specific ideology by taking advantage of social media algorithms. Social bots were found to be generating a large amount of content and possibly distorting online conversations during the 2016 U.S. presidential election. Terrorist organizations are actively using bots to promote terrorist propaganda and proselytize online extremism. As mentioned above, the evolving capability threats of bots follow other technological developments. To meet the challenge, Botometer, a supervised machine learning-based bot detection tool, has been developed by Indiana University [3]. By giving it a Twitter account, Botometer extracts over 1,000 available features from Twitter API and produces a classification bot score. The higher the score, the more likely the account is controlled by software. There are two kinds of scores based on the account's language: English and universal. The English score is suitable to use for English users. It was trained on English-based features; the universal score, on the other hand, is suitable to be used for non-English accounts. In our study, we use the English bot score. To help users understand the feature's contribution to the score, Botometer produces another six additional subscores: user meta-data, friends, content, sentiment, network, and timing, all of which exclude linguistic features.

An urgent need for accurate detection of propaganda spread by potential foreign adversaries has arisen during the COVID-19 pandemic. Consequently, in this study, we focus on fine-tuning the contextualized embeddings to detect accounts spreading propaganda and concentrate on pro-China accounts as an example.

## 3. Methodology

In this section, we describe the adopted data and methodgies applied in the study.

### 3.1 Pre-Trained Data

The main challenge in using semi-supervised learning, as mentioned above, is devising a labeled dataset. In this study, we manually collected a list of pro-China and neutral subReddit threads from trending subreddit threads. The list was reviewed by two neutral experts. Some example threads used for pro-China titles included: /r/Sino, /r/communism; examples for neutral threads included: /r/Coronavirus/, /r/technology/. Reddit API was used to collect the titles from the listed subreddit treads. In total, 15,371 subreddit titles were collected: 9,259 were labeled as neutral, 6,112 as pro-China titles. Any duplicates were removed. Table 1 gives a glimpse of what the dataset looks like.

The collected Reddit dataset was trained on RoBERTa for 1 epoch using Simple Transformers[1], a library based on the Transformers library of HuggingFace. The result is shown in Table 2. The model achieved 89.46% accuracy with a Matthews Correlation Coefficient (MCC) of 0.77749.

Our target Twitter dataset was provided on Kaggle[2]. The dataset was a collection of globally-sourced tweets containing COVID-19-related hashtags on Kaggle during the month of March, 2020. After filtering the language for English, the total dataset contained 7,459,350 tweets. In the preprocessing steps, '\n' was replaced with a space, and stop words were removed. All the emojis, hashtags, and mis-spelling features of the tweets were kept for data-mining purposes. The pre-processed tweets were fed into the trained RoBERTa model from the previous step. The prediction revealed 2,832,260 tweets labeled as neutral (labeled 0), 4,627,090 as pro-China (labeled 1). In total, there were 2,218,337 unique users in the dataset. The number of Twitter accounts producing tweets using COVID-19-related hashtags in March, 2020 had a long-tailed distribution, i.e., most of the users tweeted less than 10 times while a few extreme users tweeted about the topic up to 36,436 times.

---

[1] https://github.com/ThilinaRajapakse/simpletransformers

[2] https://www.kaggle.com/

**Table 1. Example titles in Neutral and Pro-China dataset**

|  | Label | Title |
|---|---|---|
| Neutral | 0 | 1. The president from Slovenia has decided to learn the language of every Europen country affected by Coronavirus to send them an encouraging message. Here is the British one. He also did: Spain, Italy, Poland, ... more<br>2. Sweden has closed the country's last coal-fired power station two years ahead of schedule.<br>3. Coronavirus in Russia: Anti-government protests go online |
| Pro-China | 1 | 1. Ian Goodrum is an American communist journalist currently living in China and working for China Daily. His twitter page is filled with well-sourced information debunking capitalist propaganda. Check him out for sources!<br>2. Western Hypocrisy<br>3. The Finnish Communist Revolution of 1918: An Indie Docu-series |

**Table 2. Result of Reddit dataset trained on RoBERTa**

| Accuracy | MCC | TP | TN | FP | FN | Eval loss |
|---|---|---|---|---|---|---|
| 0.89466 | 0.77749 | 255 | 433 | 38 | 43 | 0.33572 |

We used the Botometer English bot score to get a third party view on how likely an account was to be a bot account. At the time of this writing, we have collected bot scores for 17,000 accounts. We removed 1,444 accounts, 1,331 of which appear to be suspended, and 113 of which have shown id-mismatch to our dataset. At this point, combining the tweets with the Botometer score, we have a dataset consisting of 15,556 users, 27,764 labeled as neutral (labeled 0), 45,683 as pro-China (labeled 1); with 73,831 tweets in total.

### 3.2 Statistical Analysis

The distribution of the English bot score for pro-China and neutral accounts is shown in Figure 1. The x-axis is the score ranging from 0 to 1 (left to right); the higher, the more certain Botometer concludes that the user is a bot. The distribution of content, friend, network, sentiment, temporal, user bot score of pro-China (labeled 1), and neutral (labeled 0) accounts are shown in Figures 2 to 7.

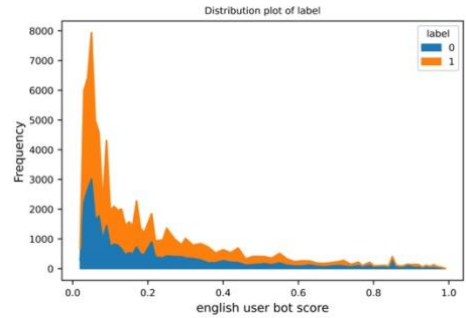

**Figure 1. The distribution of English bot score of pro-China(1) and neutral accounts(0)**

To find out if the distribution for the two scores was statistically different, we conducted a Two-sample Kolmogorov–Smirnov test. The null hypothesis was that the two distributions would be the same if the p-value was less than 0.05. To reject the null hypothesis, i.e., the two target distributions would be different. The results are shown in Table 3, which shows that on all the bot score objects, the distributions between pro-China and neutral groups were different.

**Table 3. Two-sample Kolmogorov–Smirnov test between pro-china and neutral bot score distribution**

| Two-sample Kolmogorov–Smirnov test | | |
|---|---|---|
| $H_0$ = pro-China and neutral distributions are the same | | |
| obj bot score | p-value | Reject $H_0$ |
| English | 0.0000000423 | True |
| Content | 0.0000000000 | True |
| Friend | 0.0000000000 | True |
| Network | 0.0000011932 | True |
| Sentiment | 0.0000000000 | True |
| Temporal | 0.0000000000 | True |
| User | 0.0000001471 | True |

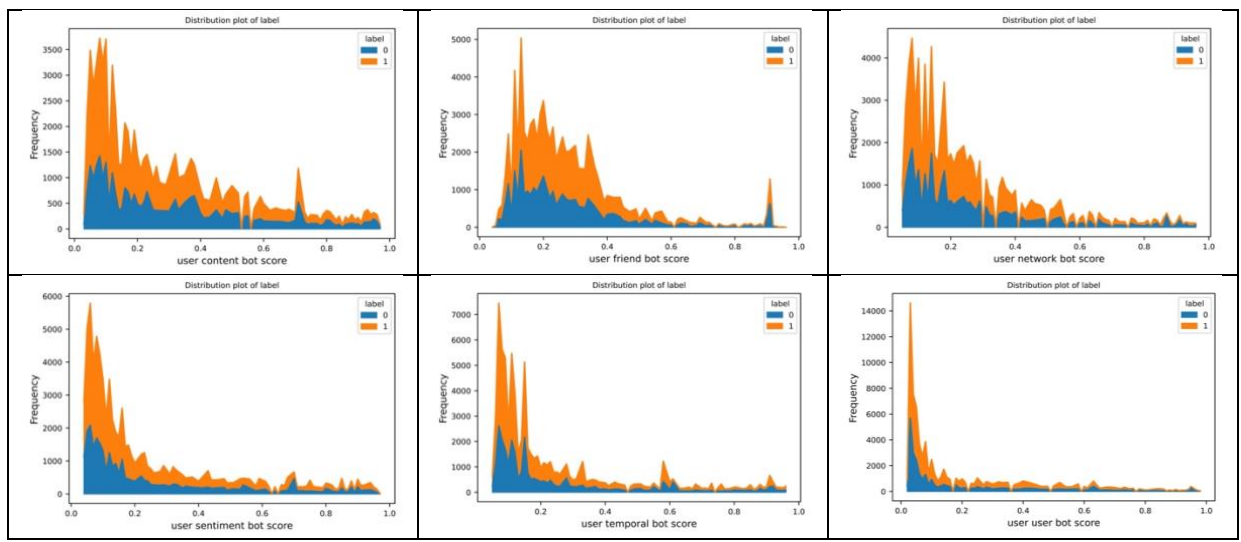

**Top: Figure 2. content bot score; Figure 3. friend bot score; Figure 4. network bot score;
Button: Figure 5. sentiment bot score; Figure 6. temporal bot score; Figure 7. user bot score**

### 3.3 N-Gram Analysis

We also conducted a distinct n-gram test on the whole twitter dataset. The result is shown in Table 4. In the distinct n-gram test, any words that showed up in both groups were dropped.

In the neutral group of the distinct 2-gram analysis: we noticed that a few left and right extreme political twitter accounts showed up, such as "@SmartDissent," "@WeAreALPA"; as well as hashtags like "#WeStandWithItaly," which is related to a petition aiming to ask the eurozone government to adjust spending to address the crisis. In the pro-China group, the distinct 2-gram analysis appears to show that they are actively tweeting much more than the neutral group. The upper bound for the neutral group had a frequency of 300, while the pro-China group had an upper bound of 35,000. The '#moneyforthepeople' and "$2000/month" are hashtags related to a petition "$2000/month to every American #moneyforthepeople #covid19"; content related to the hashtag "#todaynncoronavirus" had all been removed at the time of writing. Hashtags related to local politics were also picked up: "#PAPol" is short for Pennsylvania politics, '#TXpolitics' is short for Texas politics; both hashtags were used a lot by a bot account @openletterbot.

In the neutral group of the 3-gram analysis, the tweets appeared to be alarmed by the COVID-19 pandemic while at the same time still keeping hope and a positive attitude. As the pro-China group consisted of using a call for action tone, the words "act, must" appeared many times. In the 4-gram and 5-gram analysis, the finding aligns with the previous analysis.

Overall, the pro-China group has a much higher tweeting frequency than neutral groups. Even though bigram analysis picked up a few far-left and far-right political extremist-related accounts, a positive attitude while showing alarm about the COVID-19 situation was more in evidence. The Pro-China group used more slogan- and action-driven words. However, we did not see the conspiracy-driven words promoting violence that were discovered in other research literature [5].

### 3.4 Discussion

The importance of the detection of propaganda by foreign adversaries is indubitable. Still, the challenge remains for social media platforms to find out how to maintain freedom of speech and avoid malicious attacks from foreign entities. We aim to help social media platforms identify malicious actors by revealing the sources of the news they are broadcasting for their audience and users.

This study explored the possibility of using contextualized embeddings for propaganda detection on a cross-platform matter. We used detecting Chinese propaganda accounts for demonstration. The method may potentially be able to be applied for finding other propaganda-driven initializations.

Twitter and Stanford Internet Observatory have published their findings on account-coordinated activity in these areas: the pro-Kremlin, anti-opposition, and anti-Western content initialized by Russia [4]; the youth wing of the ruling Justice and Development Party (AKP)'s operation targeting Turkish citizens; accounts engaged in inauthentic coordinated activity to promote the Serbian Progressive Party (SNS), the party of the President of Serbia Aleksandar Vučić to attack their political opponents and to amplify content from news outlets favorable to them [4]. One of the possible drawbacks of the proposed cross-platform method we used is that the writing style may be different depending on different social media platforms [17]. However, other studies have likewise shown Chinese propaganda accounts to be publishing disinformation.

Table 4. Distinct n-gram test result. Top to bottom shows the n-gram result for 2-gram, 3-gram, 4-gram, 5-gram

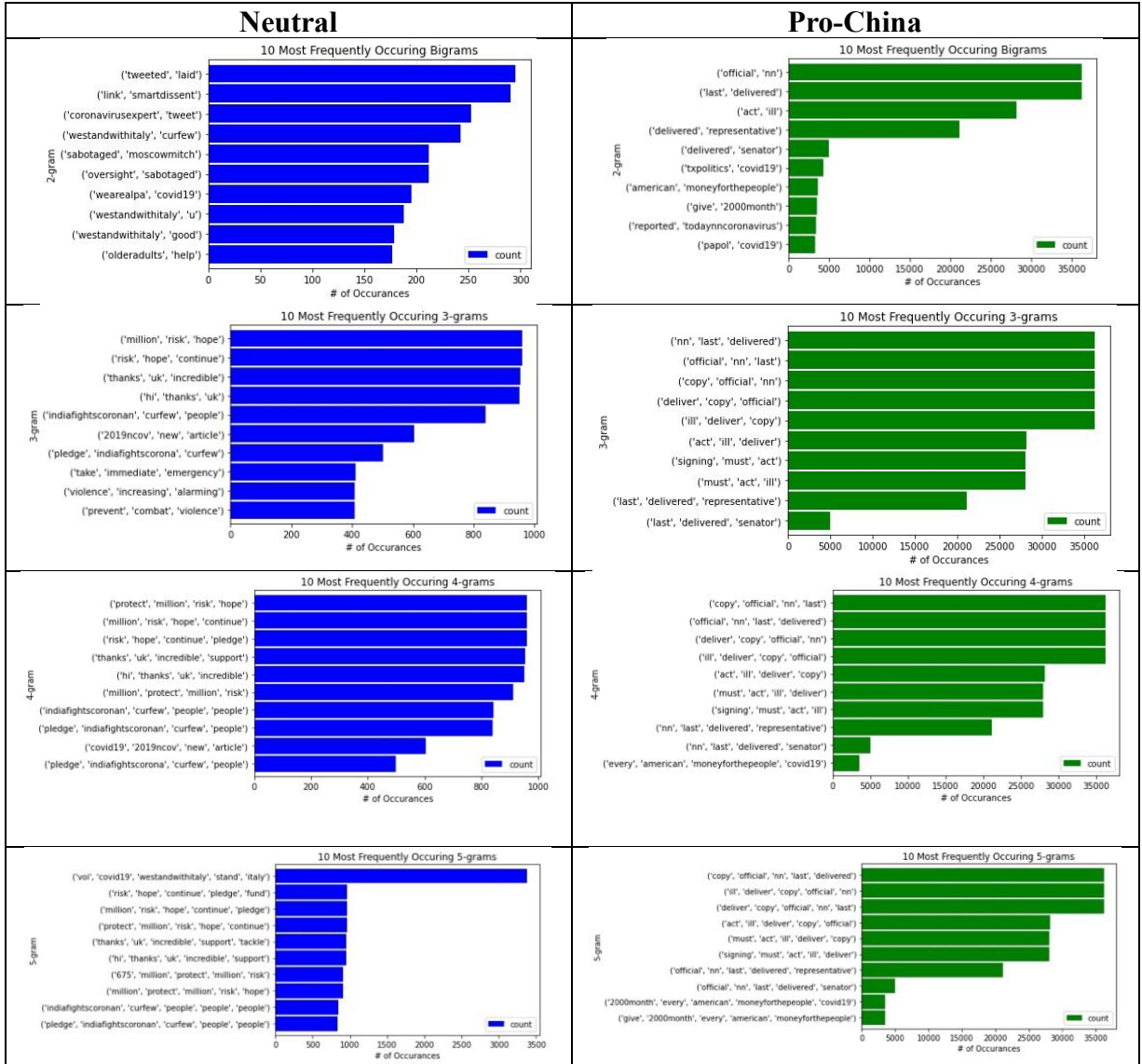

Besides the question of different cross-media writing styles, we also noticed that the n-gram analysis might be over-representing the active tweet user's voice. In the future, we might consider using normalization mechanisms on a per-user basis.

The Botometer is widely used in academia as a third-party validation for bot detection. However, some research has also exposed false positives, and there is a problem of dynamics, i.e., the botometer scores one gets may be different from time to time and hard to replicate [2].

## 4. Conclusion

To study the dissemination of information in this challenging COVID-19 pandemic period, we have used RoBERTa for propaganda detection with cross-platform features and validated the bot score with Botometer. Our findings confirm that neutral and pro-China groups use different words. In the n-gram analysis, pro-China groups appear to have more call-to-action words with local politics. Moreover, although China-related names did not show up, tweeting was at a much higher frequency. Diversely, the neutral group appeared to voice alarm and promote the attitude of maintaining hope while acknowledging the risk during the pandemic.